\documentclass[a4paper]{aa}
\usepackage{graphicx}
\usepackage{txfonts}
\graphicspath{{fig/}}
\usepackage{url}
\usepackage[english]{babel}
\usepackage{color}
\usepackage{latexsym}

\newcommand{\Hm}{\ensuremath{\mathrm{H}^-}}
\newcommand{\Qdiss}{Q_\mathrm{diss}}

\authorrunning{T.\ Heinemann et al.}
\titlerunning{Radiative transfer in decomposed domains}
\title{Radiative transfer in decomposed domains}
\author{T. Heinemann\inst{1,2}
        \and
        W. Dobler\inst{3,4}
        \and
        \AA. Nordlund \inst{2}
        \and
        A. Brandenburg\inst{1}}
\institute{NORDITA, Blegdamsvej 17, DK-2100 Copenhagen \O, Denmark,
        \email{theine@nordita.dk}
         \and
         Niels Bohr Institute, Copenhagen University,
         Juliane Maries Vej 30, DK-2100 Copenhagen \O, Denmark
         \and
         Department of Physics and Astronomy, University of Calgary,
         Calgary, Alberta, Canada
         \and
         Kiepenheuer-Institut f\"ur Sonnenphysik,
         Sch\"oneckstra\ss{}e 6, D-79104 Freiburg, Germany
}\date{\today,~ $ $Revision: 1.204 $ $}

\abstract
{}
{
An efficient algorithm for calculating radiative transfer on massively
parallel computers using domain decomposition is presented.
}
{
The integral formulation of the transfer equation is used to divide
the problem into a local but compute-intensive part for calculating
the intensity and optical depth integrals, and a nonlocal part
for communicating the intensity between adjacent processors.
}
{
The waiting time of idle processors during the nonlocal communication part
does not have a severe impact on the scaling.
The wall clock time thus scales nearly linearly with the inverse
number of processors.
}
{}

\begin{document}

\maketitle

\section{Introduction}\label{sec:I}

Over the past one or two decades tremendous advances have been made in achieving
high resolution power in computational astrophysical fluid dynamics; see Haugen
et al.\ (\cite{HBD03}) for a $1024^3$ simulation of hydromagnetic turbulence and
Kaneda et al.\ (\cite{Kan03}) for a $4096^3$ simulation without magnetic fields.
Such high resolution is now possible mainly due to the availability of massively
parallel computers allowing superb performance at a low price, especially if
off-the-shelf personal computers can be interconnected using standard Ethernet
switches. This poses no major difficulty for the usual finite difference schemes
that allow the computational domain to be decomposed into smaller sub-domains,
because the necessary communication between processors is limited to a small
neighborhood of the
processor boundaries.

Radiative transfer calculations fall generally outside this class of problems,
because the transfer equation is intrinsically nonlocal. Physically speaking,
information travels at the speed of light when the gas is optically thin. Thus,
from one time step to the next, the change in intensity in the domain of one
processor can affect the radiation field on many other processors even if they
are far apart.

In this paper we describe a simple method that renders the transfer problem
essentially local -- at least as far as the bulk of the computational cost is
concerned. By using the integral formulation of the transfer equation, the
intensity may be written as a local integral term plus an
attenuated boundary term. The local integral term may be computed
in parallel by all processors whereas only the boundary term, which
may be applied after the integrals have been computed on
all processors, requires communication between processors.

The attenuated boundary terms only need to be computed on and communicated
across those boundaries where the radiation leaves each sub-domain (hereafter
referred to as downstream boundaries). The corresponding update of the
interior of each sub-domain may be carried out afterwards and again is a
completely local operation.

The efficiency of our parallelization method depends heavily on how rapidly
the attenuated boundary terms may be obtained during the communication step.
Due to this dependency, our method is only practical if
the radiation is strictly along straight lines and does not diffuse in the
transverse direction via interpolation, as in the short characteristics
method, for example.
This is discussed in detail in Sect.~\ref{sec:IP}.

Our technique may be contrasted with other popular approaches to solving
the transfer equation in decomposed domains. In the multiple wavefront method
(Nakamoto et al.\ \cite{NUS01}), parallel efficiency is achieved by allowing
different ray directions to be treated simultaneously, trying to avoid
multiple tasks for the same processor and minimizing the number of idle
processors. This method does not impose any restrictions on the radiative
transfer scheme, but efficient parallelization requires a large number of ray
directions and frequencies to be treated.
If a non-diffusive scheme is employed, our technique
appears simpler and remains efficient even when the number of CPUs greatly
exceeds the number of ray directions and frequencies.

\section{The transfer equation}\label{sec:TE}

Radiation couples with the equations of fluid dynamics both through
radiative heating and cooling and, if the temperatures are high
enough, through radiative pressure.
For the applications that we currently have in mind (e.g.\ stellar convection
and protostellar accretion discs), only heating/cooling is important, so
$\vec{\nabla}\cdot\vec{F}$ enters the energy equation
\begin{equation}\label{eq:energy-rad}
  \rho\frac{De}{Dt}+p\vec{\nabla}\cdot\vec{u}
  +\vec{\nabla}\cdot\vec{F}=\Qdiss,
\end{equation}
where $\rho$ is the mass density, $p$ is the pressure,
$\vec{u}$ is the fluid velocity,
$e$ is the internal energy per unit mass,
$\vec{F}$ is the energy flux, and $\Qdiss$ is
the (kinetic and/or magnetic) energy dissipation.
For radiative energy transfer the energy flux is given by
\begin{equation}
  \vec{F}=\oint_{4\pi} d\Omega \int\limits_0^\infty \! d\nu \;\,
  \hat{\vec{n}}\,I_\nu(\hat{\vec{n}}) ,
\end{equation}
where $I$ is the specific intensity giving the amount of energy transported by
radiation per unit frequency range per unit area per unit time into a solid
angle $\Omega$ in the direction $\hat{\vec{n}}$.

To determine the specific intensity one has to solve the transfer equation
(e.g.\ Mihalas \& Weibel-Mihalas \cite{mihalas}),
\begin{equation}\label{eq:transfer}
  \hat{\vec{n}}\cdot\vec{\nabla}I_\nu=\chi_\nu(S_\nu-I_\nu),
\end{equation}
where $\hat{\vec{n}}$ is the unit vector in the direction of propagation,
$\chi_\nu$
is the opacity (per unit volume) or inverse mean free path of a photon, and $S$
is the source function, which gives the ratio between emission and absorption.

The transfer equation (\ref{eq:transfer}) is here written in its time
independent form. This is appropriate for the non-relativistic case, where the
maximum fluid velocity is much lower than the speed of light. The flux
divergence is then given by
\begin{equation}\label{eq:div-energy-flux}
  \vec{\nabla}\cdot\vec{F}=
  \oint_{4\pi} d\Omega \int\limits_0^\infty\! d\nu \;\,\chi_\nu(S_\nu-I_\nu) .
\end{equation}

Following Nordlund (\cite{nordlund}), we define $Q_\nu=S_\nu-I_\nu$,
giving the cooling rate per ray direction and infinitesimal
frequency interval, and the optical depth scale
$\tau_\nu=\int\!\chi_\nu\,ds$,
where $s$ is measured along the propagation direction of the ray. It is then
possible to rewrite
(\ref{eq:transfer}) as
\begin{equation}\label{eq:transfer-differential}
  \frac{d Q_\nu}{d\tau_\nu}=\frac{d S_\nu}{d\tau_\nu}-Q_\nu.
\end{equation}
This equation may also be written in integral form,
\begin{equation}\label{eq:transfer-integral}
  Q(\tau)
  = Q(\tau_0)e^{\tau_0-\tau}
    + \underbrace{\int_{\tau_0}^\tau\!e^{\tau'-\tau}\frac{dS}{d\tau'}d\tau'}
      _{Q^{\rm(intr)}(\tau)},
\end{equation}
where the explicit reference to the frequency $\nu$ has been dropped.
By using $Q$ instead of $I$ numerical precision is retained even when
the optical depth is very high and $I$ approaches $S$ very closely.

In general the source function $S$ may depend on the intensity itself, i.e.\
on a nonlocal quantity, turning (\ref{eq:transfer-integral}) into an
integro-differential equation which has to be solved by means of an iterative
scheme, such as Accelerated Lambda Iteration (see Olson et al.\ \cite{olson}).
However, during each iteration step the source function is given (i.e.\ taken
from a previous iteration step) and in the following we may -- without loss of
generality -- assume that the source function is independent of
intensity.

\section{The radiative transfer scheme}\label{sec:rts}

For the sake of simplicity we here assume that the set of ray
directions is chosen in such a way that all rays travel directly through
neighboring grid points. This will suffice to motivate our method and we
delay the discussion of interpolation schemes for solving the transfer
equation for arbitrary ray directions until Sect.~\ref{sec:IP}.

With the above assumption, it is in principle easy to solve
(\ref{eq:transfer-differential}) for all ray directions. Given the cooling
rate at a mesh point $n{-}1$, the discretization of
(\ref{eq:transfer-integral}) enables us to compute the cooling rate at the
next mesh point $n$ in the direction of the ray. Once a given boundary
condition is adopted, it is thus possible to determine the cooling rate in the
entire simulation box by stepping successively along the ray.

However, in the case of domain decomposition, only those processors adjacent
to a boundary of the simulation box are able to immediately compute the
correct cooling rate within their sub-domain. All other processors have to
wait until they are provided with boundary information from a neighboring
processor that already has determined the correct cooling rate. Without
further sophistication, this would imply that most computation related to the
radiative transfer problem is not carried out in parallel and valuable CPU
time is spent in waiting.

Fortunately, for a local source function (e.g.\ independent of
mean intensity), the integral term $Q^{\rm(intr)}$ in
(\ref{eq:transfer-integral}) represents a valid solution of the transfer
equation within each sub-domain, apart only from the contribution from
the upstream boundary.
We call this the \emph{intrinsic solution},
\begin{equation}
  Q_{n}^{\mathrm{(intr)}}=Q_{n-1}^{\mathrm{(intr)}}e^{-\delta\tau_{n-1/2}}
  +\int_{\tau_{n-1}}^{\tau_{n}}\!e^{\tau-\tau_n}\frac{dS}{d\tau}d\tau,
\end{equation}
with
\begin{equation}
Q_{0}^{\mathrm{(intr)}}=0
\quad\textrm{and}\quad
\delta\tau_{n-1/2} = \tau_n-\tau_{n-1}.
\end{equation}
The complete solution for an arbitrary boundary condition $Q_0$ may be obtained
by simply adding the correction term $Q_0\,e^{\tau_0-\tau_n}$ to the intrinsic
solution on all inner points,
\begin{equation}
  Q_n=Q_0\,e^{\tau_0-\tau_n}+Q_{n}^{\rm{(intr)}}.
\end{equation}

In order to reduce the idle time of the individual processors we split
the calculation of the cooling rate into three distinct parts, two of which
may be carried out by all processors in parallel.
How this works in detail is illustrated in the following.

\subsection{Non-periodic boundaries}

\begin{figure}[tb]
  \begin{center}
    \includegraphics[width=0.45\textwidth]{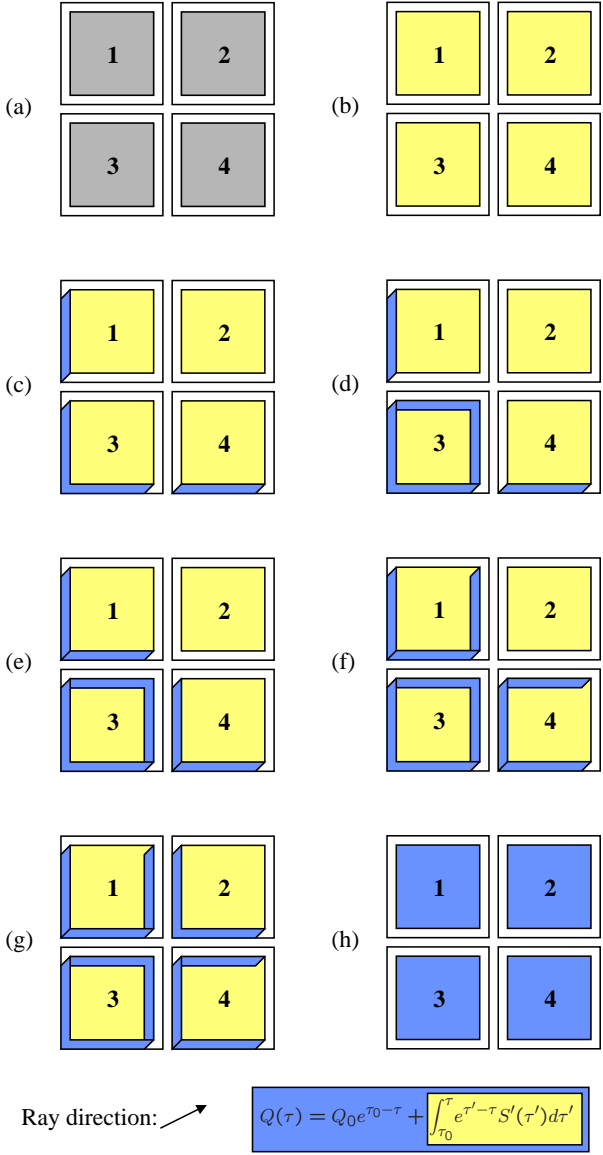}
  \end{center}
  \caption{Illustration of the radiative transfer scheme. See text for details.}
  \label{fig:boxes}
\end{figure}

We first assume that our computational domain is non-periodic in all
spatial dimensions. As far as radiative transfer is concerned, this is the
simplest case -- the periodic case
is treated in the next subsection.

The basic procedure to obtain the solution to the radiative transfer problem
in decomposed domains is illustrated in \mbox{Fig.~\ref{fig:boxes}}. For the
sake of simplicity, the computational domain is two-dimensional and divided
into four sub-domains. The transfer equation is solved for rays along the
direction indicated. The generalization to three-dimensional domains is
trivial.

In the following, $Q_0$ refers to the cooling rate on the local
upstream boundary of a given sub-domain, which either coincides with the
boundary condition of choice for the entire computational domain (global
boundary) or overlaps with the downstream boundary of a neighboring processor
in the direction opposite to the ray.

The first step is to obtain the intrinsic solution $Q^{\rm(intr)}$ within
each sub-domain, assuming a vanishing cooling rate at the boundary.
This corresponds to evaluating the integral in
Eq.~(\ref{eq:transfer-integral}).
For each point the cooling rate and the optical depth $\tau{-}\tau_0$ are
stored. This step can
be carried out by all processors in parallel since no information is
required from outside the processor (\mbox{Fig.~\ref{fig:boxes}b}).

The communication part follows next.
The local boundary cooling rate $Q_0$ in the lower ghost zone of processors 3
and 4, as well as in the left ghost zone of processors 1 and 3 are given by the
global boundary condition of choice (\mbox{Fig.~\ref{fig:boxes}c}). Since all
its upstream boundaries are set, processor 3 can immediately compute the
correct cooling rate on its upper and right boundaries by adding the correction
term $Q_0\,e^{\tau_0-\tau_N}$ to the cooling rate obtained from the intrinsic
solution (\mbox{Fig.~\ref{fig:boxes}d}). Here $\tau_0$ and $Q_0$ refer to a
point in the left (lower) ghost zone and $\tau_N$ to the corresponding point
at the upper (right) boundary along the ray.

Now that the correct cooling rate on the upper (right) boundary of processor 3
is available, this information is communicated to processor 1 (4) where the
boundary condition in the lower (left) ghost zone is set
(\mbox{Fig.~\ref{fig:boxes}e}). Likewise, processor 1 (4) is now able to
compute directly the cooling rate on its right (upper) boundary and can
send the values to processor 2 (\mbox{Figs.~\ref{fig:boxes}f and g}).

In \mbox{Fig.~\ref{fig:boxes}g} all information necessary to solve the full
transfer equation on every point on all processors is available and the
communication part is finished. The last step is again carried out by all
processors in parallel and independently of each other. It amounts to simply
adding the correction term $Q_0\,e^{\tau_0-\tau_n}$, this time on all inner
points in the sub-domain (\mbox{Fig.~\ref{fig:boxes}h}).

\subsection{Periodic boundaries}\label{sec:pb}

For many applications it is convenient to assume periodicity of the
simulation box in one or more spatial directions.
An example is convection in an infinitely extended plane-parallel layer.
While this is trivial to implement for the HD- and MHD-part of a scheme,
periodicity introduces a potential difficulty to the radiative transfer
scheme.

In the non-periodic case there is always at least one processor where all
upstream boundaries are entirely set, once the boundary condition for the whole
simulation box has been used. In the example setup of the previous subsection
this would be processor 3. By determining the cooling rate on its downstream
boundaries and communicating to all its neighbors, all upstream boundaries of
these neighbors are entirely set, and so on. This implies that each processor
has to propagate boundary values only once through its domain.

\begin{figure}[tb]
  \begin{center}
    \includegraphics[width=0.45\textwidth]{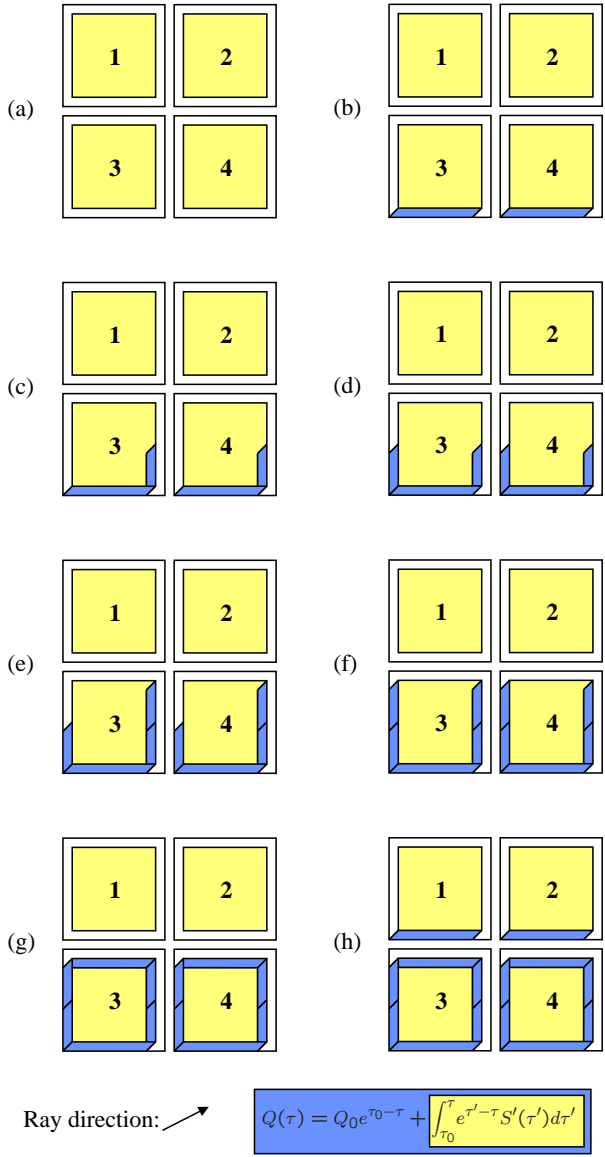}
  \end{center}
  \caption{Same as Fig.~\ref{fig:boxes}, except that here the computational
           domain is periodic in the horizontal direction.
           Note that the ray travels at an inclination of $22.5^\circ$
           relative to the horizontal axis.}
  \label{fig:boxes-peri}
\end{figure}

In contrast to the above, in the case of periodicity, it might become necessary
to propagate boundary values several times through each sub-domain.
This is illustrated in
\mbox{Fig.~\ref{fig:boxes-peri}}. The computational box in this example is taken
to be periodic in the horizontal direction, so that only the heating rates
in the lower ghost zones of processors 3 and 4 are known a priori from the
global boundary condition (\mbox{Fig.~\ref{fig:boxes-peri}}b).
Without communication, this information does not suffice to entirely
cover any of the downstream boundaries of processors 3 or 4 for the given ray
direction (\mbox{Fig.~\ref{fig:boxes-peri}}c).

In order to cover all up- and downstream boundaries of processors 3 and 4, it
is necessary to communicate the available downstream heating rates
along the periodic direction several (in this case two) times
(\mbox{Figs.~\ref{fig:boxes-peri}}d~--~\ref{fig:boxes-peri}g).

\subsection{Remarks on interpolation}\label{sec:IP}

So far we have only considered rays that pass directly through neighboring
grid points. In general we would like to take into account rays with arbitrary
inclination, so we have to interpolate in the angular
direction. According to Stone et al.\ (\cite{stone}) there are
essentially two approaches to this problem which we discuss now briefly,
taking into account their compatibility with our proposed parallelization method.

The first approach we consider is to set up exactly one ray per grid
point and trace it all the way back to the boundary of the computational
domain, interpolating the values for opacity and source function along the
way. This is usually called the \emph{method of long characteristics}. For
large numerical grids in two (three) spatial dimensions, the above
prescription is, however, hardly ever used in this form as it scales as $N^3$
($N^4$)
with the number of grid points $N$.
A common cure for this problem is to introduce in addition to the
ordinary hydrodynamical (HD) grid an arbitrarily inclined radiation grid for each ray
direction, with grid points along a number of parallel rays in that
direction (e.g.\ Nordlund
1982; Stein \& Nordlund 1988; Razoumov \& Scott 1999; Juvela \& Padoan 2005).
The values for opacity and source function are interpolated from the HD
grid onto the radiation grid, the transfer equation is solved along each
of the parallel rays, and the solution (in terms of the radiative cooling
rate) is finally interpolated back onto the HD grid.

This approach is well suited for incorporation into our
parallelization method because the radiation does not diffuse out on the
radiation grid. Hence, a ray that reaches a point on the downstream
boundary of a processor's sub-domain can be traced back to a unique point at
the upstream boundary and the attenuated boundary terms may thus be rapidly
computed during the communication step of our method. This is crucial for
keeping the idle times of the individual processors at a minimum. Furthermore,
interpolation between the two separate grids is a completely local operation
that can be carried out by all processors in parallel before and after the
communication step. Thus, the full advantage of our method can still be
exploited.
In fact, because it is a local operation, angular interpolation actually
improves our method's scaling with the number of processors since
the communication time then becomes -- relative to the overall
expense of obtaining the full solution to the transfer equation -- even less
significant.

The other approach is to use the \emph{method of short characteristics} (e.g.\ Kunasz
\& Auer \cite{kunasz88}; Auer \& Paletou \cite{paletou}; Auer et al.\
\cite{auer94}).
In this method, the rays are cut off at cell boundaries, and the radiation
intensity is interpolated onto neighboring grid points.
As a result, the radiation along rays that do not travel directly through
grid points diffuses away from the exact downstream direction.
Due to this diffusion, the radiation reaching one particular grid point on the
downstream boundary of a processor's sub-domain depends in a highly non-trivial
(unphysical) manner on the radiation coming from a substantial number of
grid points on its upstream boundary.  In order to propagate the boundary
radiation values through downstream processors one would thus essentially have
to repeat the radiative transfer solution again, but now with each processor
dependent on its upstream neighbor.  This would remove the advantage of
being able to do most of the work independently on each processor.
The short characteristics method is thus not suitable for parallelization
with the method presented here.

\subsection{Periodic rays}\label{sec:PR}

For rays traveling along a periodic direction, there is no boundary
condition to start from.
However, writing the transfer equation in its integral form,
\begin{equation}
Q_N=Q_0\,e^{\tau_0-\tau_N}+\int_{\tau_0}^{\tau_N}\!
  e^{\tau-\tau_N}\frac{dS}{d\tau}d\tau,
\end{equation}
where the subscripts $0$ and $N$ refer to corresponding points on opposite sides
of the simulation box, and using the periodicity condition $Q_N=Q_0$, it is
possible to solve for the cooling rate $Q_0$,
\begin{equation}\label{eq:periodic-correction-term}
  \left(1-e^{\tau_0-\tau_N}\right)\, Q_0
  = \int_{\tau_0}^{\tau_N}\!e^{\tau-\tau_N}\frac{dS}{d\tau}d\tau .
\end{equation}
In a decomposed domain we may thus obtain the full periodic ray solution in
much the same way as in the previous section once $Q_0$ is known. However,
additional communication is required to calculate $Q_0$ itself, which means
that we have to communicate twice through the simulation box.

\begin{figure}[tb]
  \begin{center}
    \includegraphics[width=0.48\textwidth]{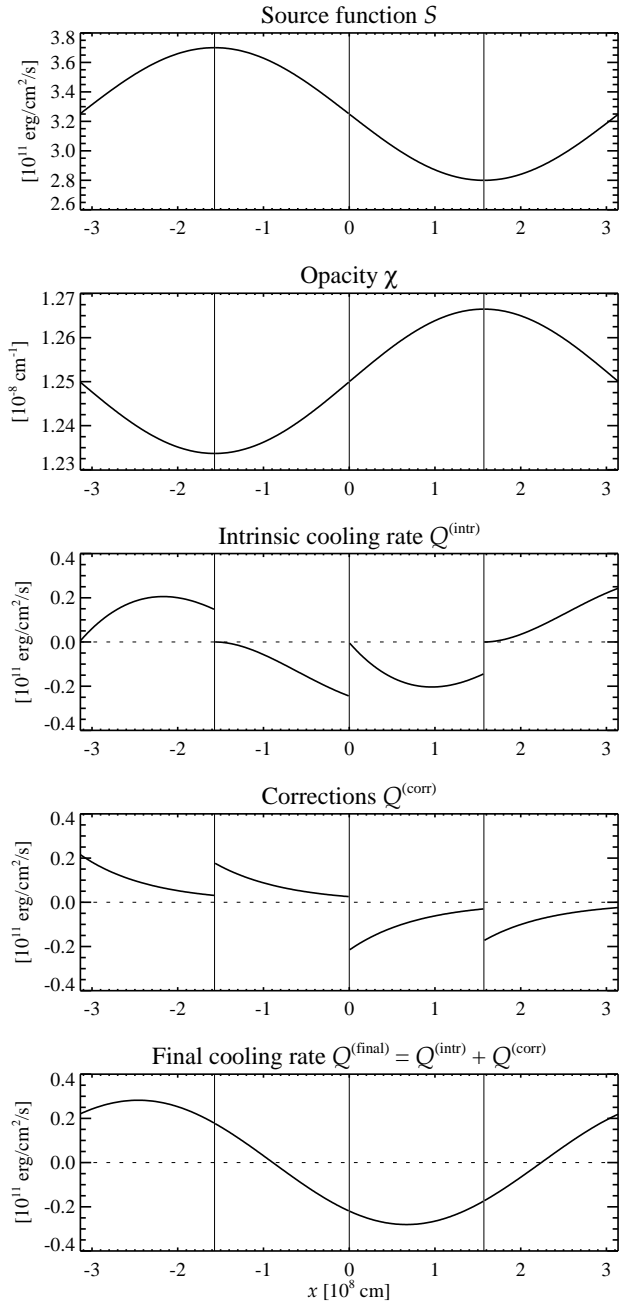}
    \caption{Illustration of the periodic ray solution for one ray traveling in
             the positive $x$-direction across 4 processors.}
    \label{fig:sinwave}
  \end{center}
\end{figure}

The periodic ray solution is illustrated in \mbox{Fig.~\ref{fig:sinwave}} for
a ray traveling in the positive $x$-direction across 4 processors. The source
function and opacity profile are depicted in the two uppermost panels,
followed by the intrinsic cooling
rate on each processor, the corresponding corrections and the final cooling
rate.

\section{Implementation into the Pencil Code}\label{sec:PC}

The method for solving radiative transfer problems, as outlined in the
previous section, has been implemented into the {\sc Pencil
Code}\footnote{\url{http://www.nordita.dk/software/pencil-code}}. This code
has been specifically designed for turbulence simulations in a parallel
computing environment using the domain decomposition scheme.

The numerical scheme consists of a third order Runge-Kutta method due to
Williamson (\cite{Wil80}) for the time stepping and sixth order centered
finite differences in space; see Brandenburg (\cite{B03}) for details. The
code is able to do domain decomposition in two spatial dimensions ($y$- and
$z$-direction) using the Message Passing Interface (MPI) for interprocessor
communications.

For the numerical solution of the transfer equation we approximate the source
function by a second order polynomial in optical depth (see Bruls et al.\
\cite{bruls}). The integral in Eq.~(\ref{eq:transfer-integral}) may then be
solved exactly. Numerical details are given in Appendix~\ref{app:ND},
available at the CDS. It is
however worth noting here that the intrinsic solution where an arbitrary but
definite boundary condition is employed may equally well be obtained by
virtue of Feautrier's (see Mihalas \cite{mihalas78}) or any other suitable
method.

In fact, we have found in another context
that on most (but not all) CPUs, the Feautrier
method is faster by up to about a factor of two, relative to the most optimized
integral method (see Appendix~\ref{app:ND} for details).
However, since the integral
method is more general (applicable for example also in cases with a combination
of Doppler effect and polarization in spectral lines), we choose to present
its implementation here.

When solving the intrinsic part of the transfer equation we store the
difference in optical depth between all grid points and the upstream boundary
in a 3-dimensional array. This allows us to quickly
compute the attenuated boundary terms and add them to the intrinsic cooling
rate on the downstream boundary (during the communication step) as well as to
all non-boundary grid-points (afterwards).

\section{Benchmarks results}\label{sec:BR}

The scaling of our method with the number of processors depends on
how many processors that can simultaneously compute the attenuated cooling rates
on the downstream boundaries of each processor.
This in turn depends on
\begin{itemize}
\item the number of ray directions.
\item the type of boundary condition (periodic or non-periodic),
\item the shape (in terms of grid points) and distribution of
      sub-domains
\end{itemize}

To investigate the above dependencies we have performed a series of benchmarks
on the Beowulf cluster at the Danish Center for Scientific Computing.  All our
tests ran on a sub-network with 302 Intel Pentium 4 machines with 3.2 GHz CPUs
and 1 GB memory, connected with Gigabit Ethernet and 10-Gigabit
uplinks. On the software side, we used the Intel Fortran 8.1 compiler
and the LAM MPI implementation.

In the benchmark series the number of processors, $N$, increases from 2 to 128 in powers
of two.  To be representative of typical applications of our
scheme, each benchmark is a short-lived 3-D hydrodynamical simulation of a
(gray) solar atmosphere near the surface.
As is explained in Appendix~\ref{app:SA}, available at the CDS,
the ionization fraction that enters
the equation of state is calculated in an iterative fashion from the
thermodynamic variables. We use the ionization fraction to calculate the
number density of negative hydrogen ions, which are the only source of opacity
in these simulations.

To find out whether the choice of boundary conditions for the radiative
cooling rate has a significant influence on the scaling, the
entire computational domain is either periodic or non-periodic in the
horizontal $x$- and $y$-directions. The boundary condition in the
$z$-direction is in both cases non-periodic.

Furthermore, we have taken into account two different sub-domain shapes,
characterized by the number of grid points in each spatial direction:
``planar'' sub-domains with $m_x\!=\!m_y\!=\!64$ grid points in the $x$- and
$y$-directions and $m_z\!=\!32$ grid points in the $z$-direction, and
``columnar'' sub-domains with $m_x\!=\!m_z\!=64$ grid points in the $x$- and
$z$-directions and $m_y\!=\!32$ grid points in the $y$-direction.  Depending on
their shape, these sub-domains are distributed in the following way. If $N$
is an even power of two, there are as many processors in the $y$-direction as
there are in the $z$-direction ($N_y\!=\!N_z$). If $N$ is an odd power of two,
then for planar domains there are twice as many processors in the
$z$-direction as there are in the $y$-direction ($N_z\!=\!2 N_y$), and for
columnar domains it is the other way around ($N_y\!=\!2 N_z$). This is
illustrated in Table~\ref{tab:sub-domains} for up to eight processors.

\begin{table}[tb]
\caption{Distribution of sub-domains in $y$ and $z$ for up to 8 processors.}
\label{tab:sub-domains}
\centering
\begin{tabular}[t]{ l c c c c }
\hline\hline
    \parbox[c]{1.7cm}{$n_{\mathrm{CPU}}$}
  & \parbox[c]{1.1cm}{\centering 1}
  & \parbox[c]{1.1cm}{\centering 2}
  & \parbox[c]{1.1cm}{\centering 4}
  & \parbox[c]{1.1cm}{\centering 8} \\
\hline
  & & & & \\
    \parbox[c]{1.7cm}{planar\\sub-domains}
  & \parbox[c]{1.1cm}{\centering\includegraphics[scale=0.5]{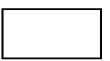}}
  & \parbox[c]{1.1cm}{\centering\includegraphics[scale=0.5]{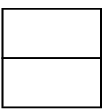}}
  & \parbox[c]{1.1cm}{\centering\includegraphics[scale=0.5]{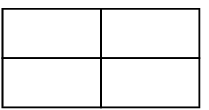}}
  & \parbox[c]{1.1cm}{\centering\includegraphics[scale=0.5]{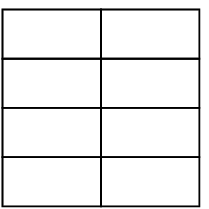}} \\
  & & & & \\
\hline
  & & & & \\
    \parbox[c]{1.7cm}{columnar\\sub-domains}
  & \parbox[c]{1.1cm}{\centering\includegraphics[scale=0.5]{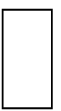}}
  & \parbox[c]{1.1cm}{\centering\includegraphics[scale=0.5]{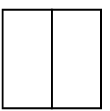}}
  & \parbox[c]{1.1cm}{\centering\includegraphics[scale=0.5]{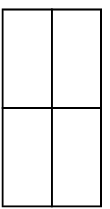}}
  & \parbox[c]{1.1cm}{\centering\includegraphics[scale=0.5]{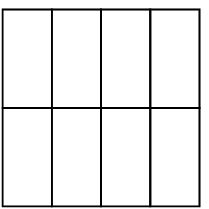}} \\
  & & & & \\
\hline
\end{tabular}
\end{table}

In our present implementation where we only solve for rays that travel through
grid points, there are three qualitatively different sets of ray directions
determined by the geometry of a grid cell:
\begin{itemize}
\item Rays along coordinate axes (6 ray directions).
\item This plus rays along face diagonals
      ($6\!+\!12\!-\!4\!=\!14$ ray directions).
\item This plus rays along the space diagonals
      ($14\!+\!8\!=\!22$ ray directions).
\end{itemize}
\noindent
Rays along face diagonals only make up $12\!-\!4\!=\!8$ additional ray
directions since rays along those face diagonals that lie in the horizontal
plane are impractical to solve for if periodic boundary conditions are
employed, and, for comparison, we left them out in the case of
non-periodic boundary conditions as well.

The reason we carried out separate benchmarks for each of the above
sets is that one could expect a decrease in scaling performance as one goes
from 6 ray directions to 22. For rays along the edges of a grid cell there is
always a whole layer of processors that can simultaneously receive the
boundary cooling rate and propagate the attenuated cooling rate computed at
the downstream boundary to the next layer. For inclined rays however, the
number of processors that can do the communication simultaneously varies as
one communicates through the entire domain and is generally less or equal than
for rays along the edges. For comparison with ordinary hydrodynamical domain
decomposition, we also included a benchmark with no radiative transfer at all
for each type of boundary condition and sub-domain shape. In total, there are
thus $4\!\times\!2\!\times\!2\!=\!16$ benchmarks per processor number.

\begin{figure}[tb]
  \begin{center}
    \includegraphics[width=0.48\textwidth]{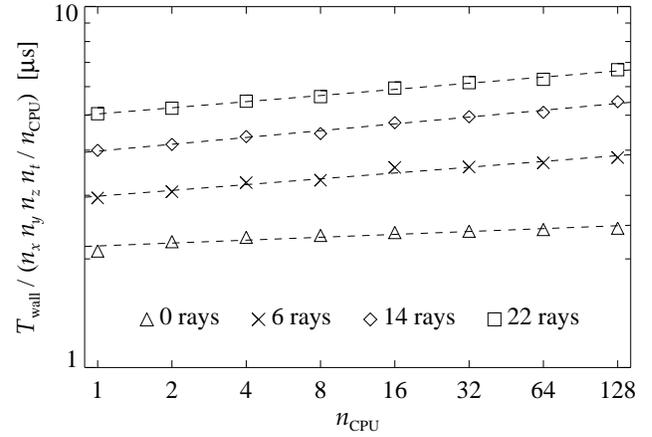}
    \caption{Wall clock time per grid point and time step in a 3-dimensional
             simulation of the solar surface for different sets of ray
             directions -- in this case for columnar sub-domains and
             non-periodic boundary conditions for the cooling rate.
             The straight lines denote least square fits.
             Perfect scaling would be represented
             by a horizontal line.}
    \label{fig:benchmarks}
  \end{center}
\end{figure}

\begin{figure}[tb]
  \begin{center}
    \includegraphics[width=0.48\textwidth]{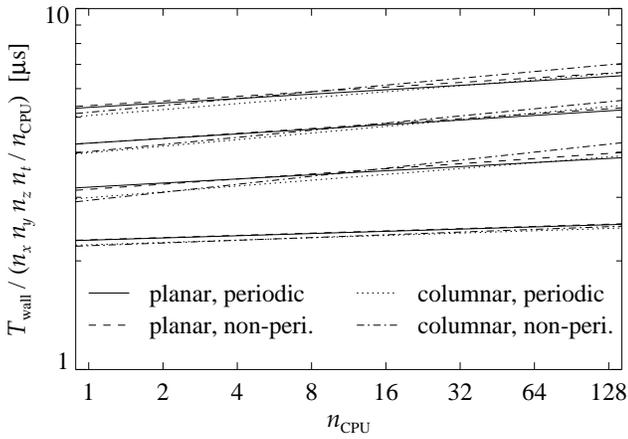}
    \caption{Least square fits to the wall clock time
             per grid point and time step
             for all sets of ray directions, boundary conditions for the
             cooling rate, and sub-domain shapes.}
  \label{fig:benchmarks-combined}
  \end{center}
\end{figure}

In Fig.~\ref{fig:benchmarks} we show the scaling of the wall clock time
per grid point and time step in a 3-dimensional simulation of the solar
surface for different sets of ray directions using columnar sub-domains
and non-periodic boundary conditions for the cooling rate.
Perfect scaling would be represented by a horizontal line in this figure.
Due to the locality of the
Navier-Stokes equations, the purely hydrodynamic benchmark series (0 rays)
are close to being perfect in the above sense.
In comparison, the radiative benchmarks (6, 14, and 22 rays) show
for each type of boundary condition and sub-domain shape
a slight increase in the wall clock time with the number of processors,
but generally with at most about 30\% as $n_\mathrm{CPU}$ varies from 1 to 128.

In Fig.~\ref{fig:benchmarks-combined} we compare the scaling behavior
for different shapes of the sub-domains, both for periodic and
non-periodic rays.
The main conclusion to be drawn from this is that the difference in
performance is surprisingly small.
Nevertheless, as a general trend one can say that the scaling is
slightly better for planar than for columnar sub-domains.
On the other hand, columnar sub-domains perform slightly better for
a small number of processors ($n_\mathrm{CPU}\le8$).
Within the accuracy of the measurements, which is not perfect due to
constantly varying network and I/O performance on the cluster in the course of
the benchmark series, we can say that neither the number of ray directions,
nor the choice of boundary conditions and sub-domains shape has an appreciable
impact on scaling with the number of processors.

\section{Conclusions}\label{sec:C}

In this paper we have presented a parallelization method for carrying out
hydrodynamical radiative transfer calculations on massively parallel computers
using the domain decomposition scheme. The proposed method is
conceptually simple and straightforward to implement. It is also flexible in
the sense that the solution of the transfer equation is not limited to the
integral method but may also be obtained by direct discretization (Feautrier
type methods).

We find that the proposed parallelization method scales almost linearly with
inverse number of processors, irrespective of the choice of boundary
conditions, sub-domain shape, or number of ray directions.
The method is thus ideal for carrying out large hydrodynamical simulations
on massively parallel supercomputers using the domain decomposition scheme.

The present implementation into the {\sc Pencil Code}
only represents a first proof
of concept. The inclusion of more ray directions, a non-uniform vertical mesh,
non-gray radiative transfer, radiation pressure, or scattering opacities are
all examples of extensions that are still possible within the framework of our
parallelization method.

Our focus has not been to optimize the intrinsic part of the radiative
transfer calculations, but already in the implementation
used for the benchmark series in Sect.~\ref{sec:BR}, one can afford 22 rays
per mesh point at a cost of about 40\% of the total time to advance
the MHD-equations (60\% relative to advancing only the HD-equations).
With the fully optimized integral method (storing and reusing exponentials
whenever possible), or with the
Feautrier method, one can afford 2 to 4 times
more rays per point, depending on the CPU and the type of network.

The number of possible applications of our method is large. It has been tested
successfully for simulations of the solar atmosphere where
even a gray opacity treatment with a small number of rays already gives
quite useful results
(but a more accurate model requires an opacity bin coverage,
cf.\ Stein \& Nordlund \cite{SN89}).
To give a list of other applications that is by no means exhaustive, the
method is directly applicable to local simulations of
accretions discs using the shearing sheet approximations, to global models
of stars or discs that are embedded in a Cartesian domain (e.g.~Dobler et
al.~\cite{DSB06}, Freytag et al.~\cite{FSD02}), as well as to studies of
radiatively driven ionization in HII regions.  A generalization to time
dependent radiative transfer would allow applications of the method
to studies of the reionization of the Universe, as well as to other
contexts where effects of the finite speed of light are important.

\begin{acknowledgements}
The work of {\AA}N was supported by grant number 21-01-0557 from the
Danish Research Council for Nature and Universe (FNU). Computing time
was provided by the Danish Sci\-en\-ti\-fic Computing Center (DCSC).
\end{acknowledgements}

\Online

\appendix

\section{Numerical details}\label{app:ND}

In order to solve Eq.~(\ref{eq:transfer-integral}) one uses polynomial
approximations for the opacity $\chi$ and the source function $S$.
In the current implementation for the {\sc Pencil Code}, the
opacity is assumed to vary linearly from a mesh point $n$ to the
next mesh point $n{+}1$ that lies in the direction of the ray, so
 the difference in
optical depth between these points, calculated by the trapezoidal
rule, is
\begin{equation}
  \delta\tau_{n+1/2}
  ={\textstyle{1\over2}}(\chi_n+\chi_{n+1})\;\delta s_{n+1/2},
\end{equation}
where $\delta s_{n+1/2}$ is the spatial distance between $n$ and $n{+}1$.
A more accurate (but also more expensive) method is based on a cubic
spline fit, using the logarithmic derivative of the opacity,
\begin{equation}\label{spline-dtau}
  \delta\tau_{n+1/2}=\textstyle{1\over2}
     [\chi_n    (1+\textstyle{1\over6} d_n     \delta s ) +
      \chi_{n+1}(1-\textstyle{1\over6} d_{n+1} \delta s ) ]\;\delta s,
\end{equation}
where
\begin{equation}
   d_n  = \left(\frac{d\ln \chi}{ds}\right)_n ,
\end{equation}
and, for brevity, we write $\delta s$ instead of $\delta s_{n+1/2}$.
While theoretically this expression may yield negative values for
$\delta\tau$, for reasonably
resolved temperature variations $\delta s$ should not exceed unity
(and certainly must not exceed 6), so in practice this does not happen.
A third alternative is
\begin{equation}
d\tau_{n} = \frac{\chi_{n+1}-\chi_{n}}{\ln(\chi_{n+1}/\chi_{n})}ds_{n} ,
\end{equation}
but in practice this turns out to be less accurate than Eq.~(\ref{spline-dtau})
above.

To calculate the cooling rate $Q^\uparrow_n$ due to an `upwards' directed
ray, between $\tau_{n}$ and $\tau_{n+1}$ the quadratic Taylor
approximation for the source function is used, i.e.~we ignore the third and
higher derivatives of $S(\tau)$.
Equation~(\ref{eq:transfer-integral}) then gives
\begin{equation}\label{eq:cooling-iteration}
  Q^\uparrow_{n} = a_{n-1/2}\, Q_{n-1} + b_{n-1/2}\, S'_{n-1} + c_{n-1/2}\, S''_{n-1}
\end{equation}
with three coefficients
\begin{equation}
  a_{n-1/2} = e^{-\delta\tau_{n-1/2}},\quad
  b_{n-1/2} = 1 - e^{-\delta\tau_{n-1/2}},
\end{equation}
and
\begin{equation}
  c_{n-1/2} = e^{-\delta\tau_{n-1/2}}(1+\delta\tau_{n-1/2}) - 1;
\end{equation}
similarly, the `downwards' directed rays give
\begin{equation}\label{eq:cooling-iteration-down}
  Q^\downarrow_{n} = a_{n+1/2}\, Q_{n+1} - b_{n+1/2}\, S'_{n+1} + c_{n+1/2}\, S''_{n+1} .
\end{equation}

One observes that in the limit of large optical depth,
$(Q^\uparrow + Q^\downarrow)/2 \rightarrow -S''$ (diffusion limit).
This demonstrates that the second derivative $S''$ needs to be taken
into account, as otherwise the numerically obtained total heating rates
would be wrong (they would still be $\propto S''$ because up- and
down-stream contributions do not cancel exactly, but with an incorrect and
resolution-dependent coefficient).

We point out that a factor of nearly two in computational speed
may be gained in the radiative transfer part of the computations,
at the expense of some storage space, by storing and reusing the
$e^{-\delta\tau_{n-1/2}}$ factors between two rays in opposite
directions. The speed increase was verified in smaller test cases,
but the timings presented in \mbox{Fig.~\ref{fig:benchmarks}} were obtained
without implementing this.

For small values of $\delta\tau$, the expressions for the coefficients
$a_n$, $b_n$, and $c_n$, must be
computed in double precision to avoid loss of precision, but
the final coefficients may be stored in single precision without
noticeable loss of accuracy.
On some CPUs further speed-up may be obtained by conditionally using
asymptotic expressions for these coefficients, while in other cases,
especially where compiler options or special libraries are available
to enable vectorization or other optimization of exponentials,
it may be faster to retain the explicit computation of exponentials.
We have implemented coding that automatically chooses between these
two alternatives, based on an initial comparison of the speeds.

The derivatives of $S$ with respect to the optical depth $\tau$
have to be computed on an irregularly spaced grid, i.e.\
\begin{equation}
  S'_n = \left.
         \left(
           \frac{\delta S_{n-1/2}}{\delta\tau_{n-1/2}}
           \frac{\delta\tau_{n+1/2}}{2}
          +\frac{\delta S_{n+1/2}}{\delta\tau_{n+1/2}}
           \frac{\delta\tau_{n-1/2}}{2}
         \right) \right/
        \overline{\delta\tau_n},
\end{equation}
\begin{equation}
  S''_n=\left.\left(\frac{\delta S_{n+1/2}}{\delta\tau_{n+1/2}}
                -\frac{\delta S_{n-1/2}}{\delta\tau_{n-1/2}}\right)
        \right/\overline{\delta\tau_n} ,
\end{equation}
where $\overline{\delta\tau_n} = (\delta\tau_{n-1/2}+\delta\tau_{n+1/2})/2$
(see Bruls et al.\ \cite{bruls}).
The procedure to compute radiative energy transport in Car\-te\-sian geometry is
then straightforward. For each ray direction there are in three dimensions --
dependent on the inclination -- either one, two or three upstream boundaries
where a certain boundary condition is employed and the iteration as defined by
(\ref{eq:cooling-iteration}) starts. By moving stepwise through the entire box
it is possible to determine the cooling rate on every single mesh point and for
all desired ray directions.

\section{Simulations of the solar atmosphere}\label{app:SA}

The benchmark we used in Sect.~\ref{sec:PC} is a short-lived simulation of
the solar atmosphere near the surface without magnetic fields. For all
benchmarks in the series (see Sect.~\ref{sec:BR} the physical
size of the simulation box is $6~\rm{Mm} \times 6~\rm{Mm} \times 6~\rm{Mm}$
in $x$, $y$, and $z$ respectively.
To account for the varying resolution of the numerical grid during the
benchmark series, the viscosity is appropriately adjusted.
The simulation box is periodic in both $x$ and $y$, even if
the boundary condition for the radiative cooling rate is
non-periodic as it is the case for half of the benchmark runs.

Furthermore, we use an equation of state that accounts for hydrogen ionization
but ignores the negative hydrogen ion, $\Hm$, and hydrogen molecule formation.
The hydrogen ionization fraction is obtained iteratively from entropy and
density by solving Saha's equation. The $\Hm$ opacity is then
calculated from the number density of $\Hm$, which is again calculated from
the number density of electrons and neutral hydrogen. The number density of
$\Hm$ is very small, so even though $\Hm$ is the most important (and, in this
simulation, only) contributor to the opacity, ignoring it in the equation of
state is justified.

\end{document}